\documentclass[twocolumn]{revtex4}

\usepackage[pdftex]{graphicx}

\begin{document}

\title{Air shower registration algorithm and mathematical processing of showers with radio signal at the Yakutsk array}


\author{I. Petrov}
\author{S. Knurenko}
\author{Z. Petrov}

\affiliation{Yu. G. Shafer Institute of Cosmophysical Research and Aeronomy, Russia.}

\email{igor.petrov@ikfia.sbras.ru}

\begin{abstract}
The paper describes the techniques and method of registration of air shower radio emission at the Yakutsk array of extensive air showers at a frequency of 32 MHz. At this stage, emission registration involves two set of antennas, the distance between them is 500m. One set involves 8 antennas, second – 4 antennas. The antennas are perpendicularly crossed dipoles with radiation pattern North – South, West – East and raised 1.5 m above the ground. Each set of antennas connected to an industrial PC. The registration requires one of two triggers. First trigger are generated by scintillation detectors of Yakutsk array. Scintillation detectors cover area of 12 km$^2$ and registers air showers with energy more than 10$^{17}$ eV. The second trigger is generated by Small Cherenkov Array that covers area of 1 km$^2$ and registers air showers with energy 10$^{15}$ – 5$\cdot$10$^{17}$ eV. Small Cherenkov Array is part of Yakutsk array and involve Cherenkov detectors located at a distance of 50, 100, 250 m. For further selection we are using an additional criterion – the radio pulse must be localized in the area corresponding to the delay time on first and second triggers. In addition, descriptions of the algorithm and the flowcharts of the program for the air shower selection and further analysis are given. This method registers EAS radio emission with energy 10$^{16}$ – 10$^{19}$ eV. With the absolute calibration, the amplitudes of all antennas converted to a single value. Air shower radio emission dependences from zenith angle and shower energy are plotted.
\end{abstract}


\maketitle

\section{Introduction}

Yakutsk array recently conducted new series of measurements of air showers radio emission, which is based on a new digital measurement technology of radio emission and dual dipole antennas oriented in the direction W-E and N-S. registration is carried out at the frequency of 32 MHz. Frequency range of 28 – 40 MHz is observed to be less influenced by noise [1]. The goal of our new measurements is not only to study nature of extensive air showers (EAS) radio emission, but obtaining an independent estimation of such characteristics as energy of primary particle and to explain the connection between longitudinal development of EAS and shape of spatial distribution of the radio emission.

\section{Short Description of Radio Array}

\begin{figure}[t]
  \centering
  \includegraphics[width=0.4\textwidth]{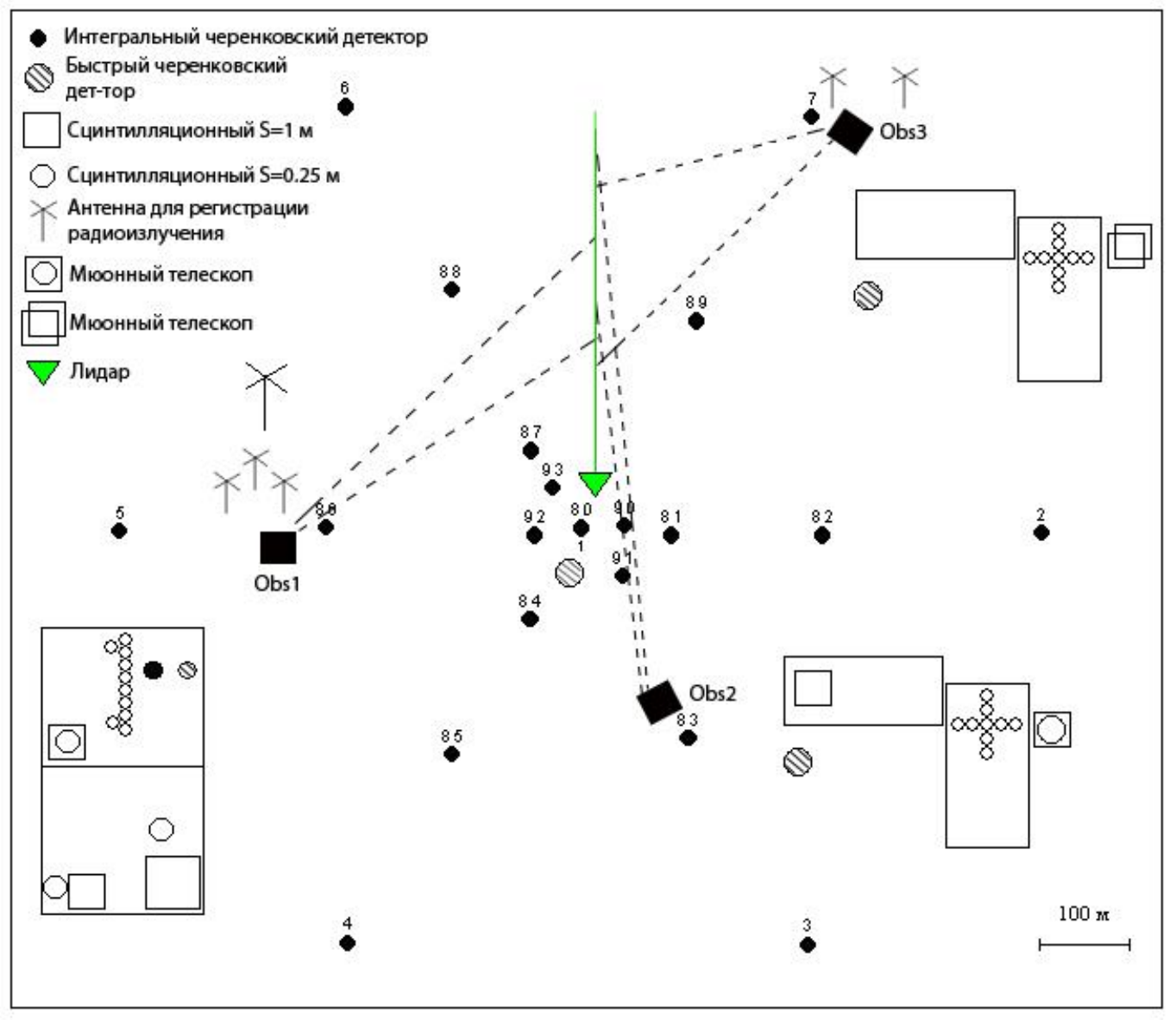}
  \caption{Arrangement of observation stations on the Small Cherenkov array.}
  \label{icrc2013-0182-01}
 \end{figure}

The array currently consists two set of antennas, distance between them is 500 m. First set consists 8 antennas, second – 4 antennas.  The antenna is a perpendicularly crossed dipoles, one to direction W-E, other N-S and lifted to the height 1.5 m above the ground. Each set is connected to an industrial computer. Directly under the antennas are electronic devices and matching amplifiers and shielding grid on the ground. Testing and calibration of measurement including the antennas is done once in a measurements season using a broadband high-frequency generator and a control PC.
Registration of radio emission is triggered by one of two event triggers (“masters”) from Yakutsk array. Location of antenna sets is shown in Fig. $\ref{icrc2013-0182-01}$.

\section {Registration Algorithm and Data Analysis}

Air shower events is triggered by one of two triggers (“master” signal). The first is the main Yakutsk array trigger, registers showers in area 12 km$^{2}$ with energy more than 10$^{17}$ eV. The second – Small Cherenkov array, registers showers in area 1 km2 with energy 10$^{15}$ – 5$\cdot$10$^{17}$ eV.

\begin{figure}[t]
  \centering
  \includegraphics[width=0.4\textwidth]{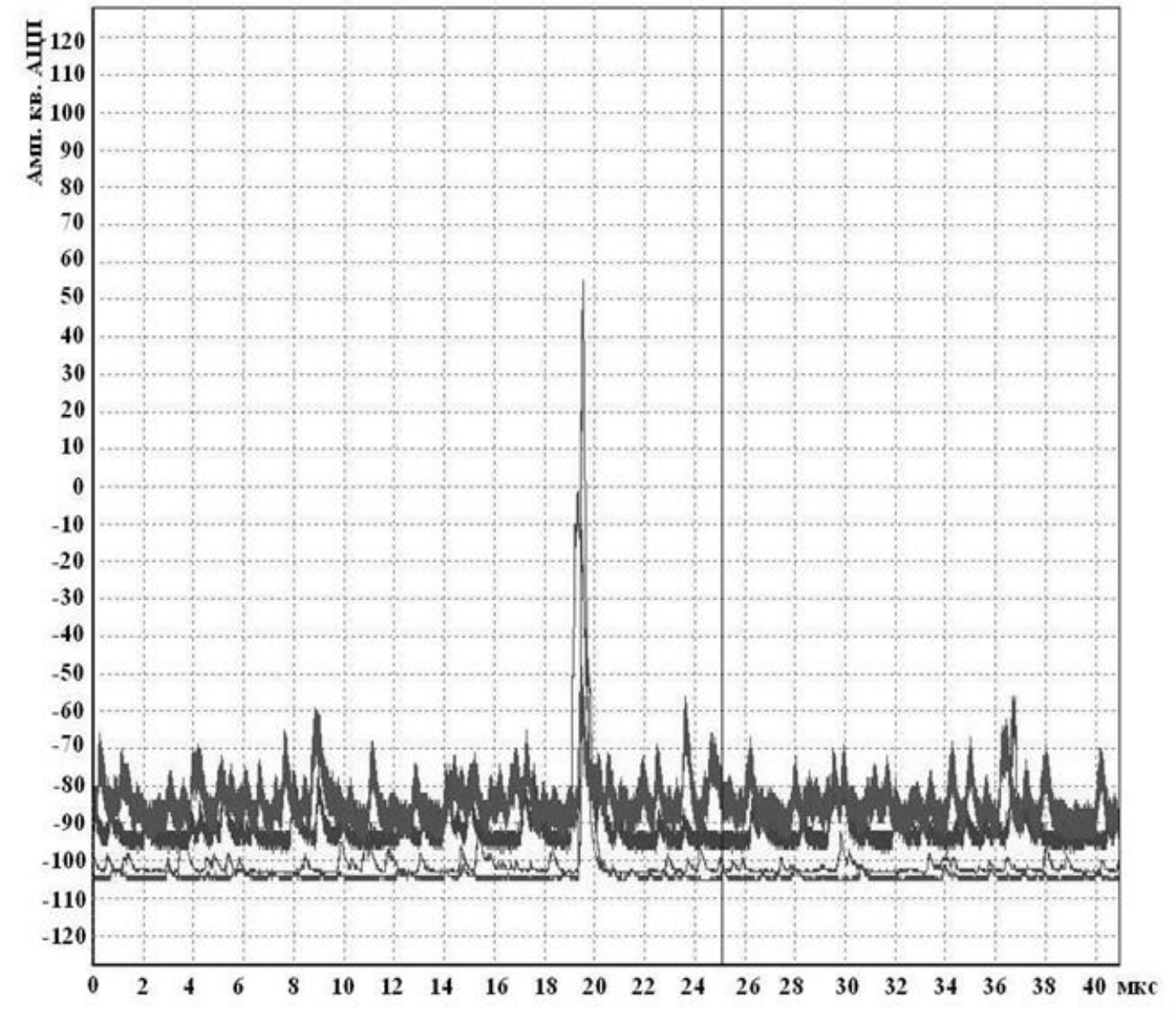}
  \caption{Radio emission pulses from 4 antennas. OX axis – 40 $\mu$s time scale, OY axis – signal amplitude in ADC samples.  Vertical line is the moment of the trigger signal arrival. }
  \label{icrc2013-0182-02}
 \end{figure}

EAS radio emission registers in the case if one of two trigger signals has been arrived. This trigger signal is the control signal for the registration of radio pulses received from antenna cable to the ADC. The data collection system is based on an industrial computer and is able to recording signal synchronously from multiple antennas simultaneously. We used fast 8-bit ADC, which provided recording prehistory (before the trigger arrival) for 25 ms and history (after the arrival of the trigger) for 15 ms (Fig. $\ref{icrc2013-0182-02}$).

\begin{figure}[t]
  \centering
  \includegraphics[width=0.4\textwidth]{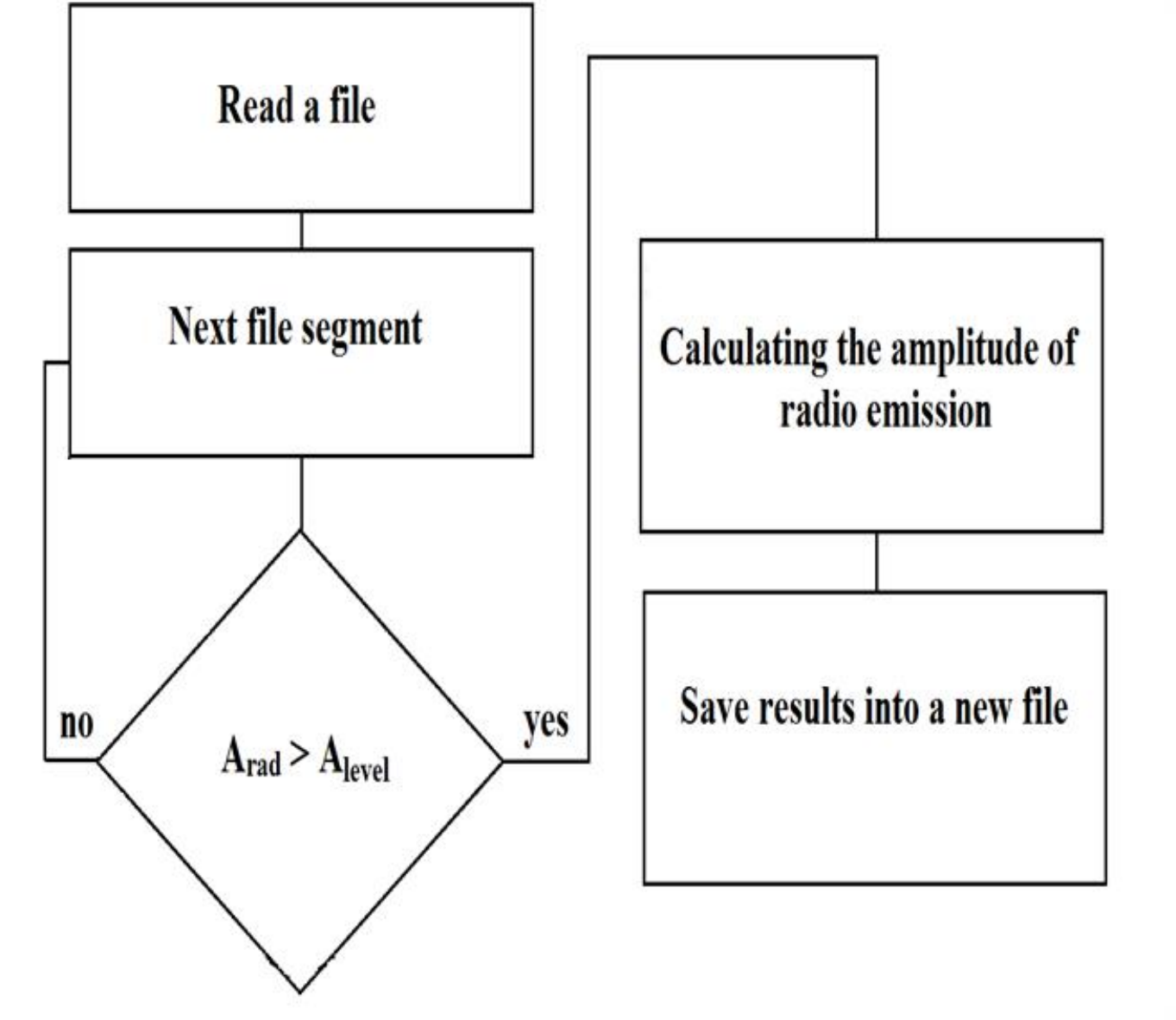}
  \caption{Radio pulse searching algorithm. A$_{rad}$ – radio pulse amplitude. }
  \label{icrc2013-0182-03}
 \end{figure}

For  pre-selection we used a program with algorithm shown in Fig. $\ref{icrc2013-0182-03}$. The program identifies EAS event if signal pulse is five times higher than noise level and pulse localized within time interval equal to “master” signal delay from Small and Large arrays. From selected data we derive pulse amplitude from antennas oriented N-S and W-E and average amplitude by formula:

\begin{equation}
        \varepsilon =
        \pm\sqrt{\left|\frac{1}{N_{Pairs}}
        \sum\limits_{i=1}^{N-1}
        \sum\limits_{j>i}^{N}
        s_{i}[t]s_{j}[t]\right|}
\end{equation}
here N$_{Pairs}$ – number of antenna pairings

\begin{figure}[t]
  \centering
  \includegraphics[width=0.4\textwidth]{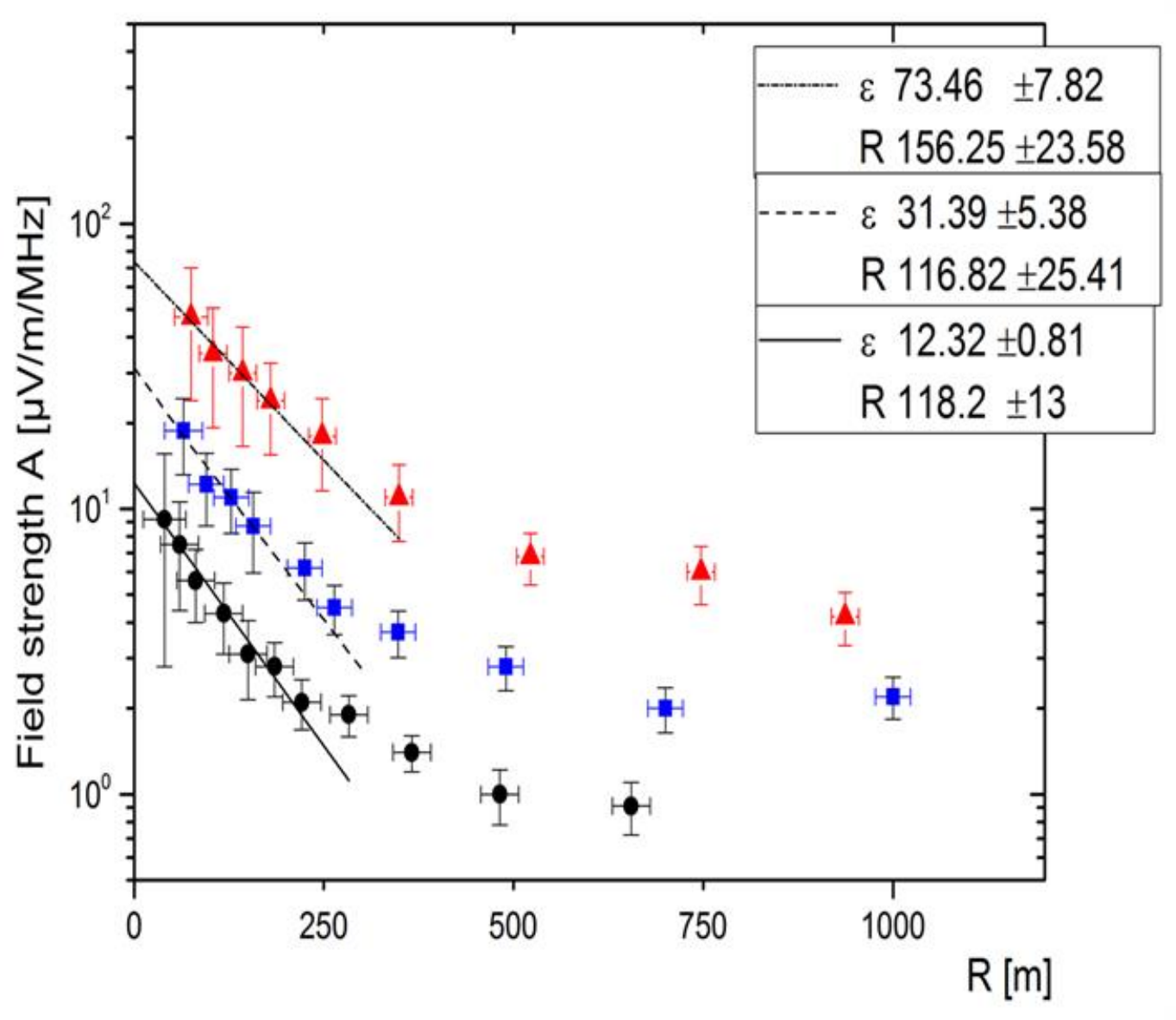}
  \caption{Average lateral distribution function of radio emission at frequency 32 MHz in showers with energy
  1.73$\cdot$10$^{17}$ eV, 4.38$\cdot$10$^{17}$ eV and 1.32$\cdot$10$^{18}$ eV. }
  \label{icrc2013-0182-04}
 \end{figure}

Other air shower characteristics are derived from data of scintillation, muon and Cherenkov detectors of main array. All data are stored in array main server and formed the basis of air shower radio emission database. The developed software for database maintenance allowed to conduct any analysis of the experimental data. As an example in Fig. $\ref{icrc2013-0182-04}$ lateral distribution function for three energies is shown.
LDF approximation is given by formula (2):

\begin{equation}
  \varepsilon(R)=
  \varepsilon_{0}\cdot\left(-\frac{R}{R_{0}}\right)
\end{equation}

The contribution of noise in measured signal can be set only by simulation measurements as carried out in [2] for LOPES data. According [2], noise impact to signal depends on a distance, in which the signal was measured. For example, at small distances, where the signal is large impact of noise is smaller than at large distances where is measured signal similar to noise level. Radio emission measurements modelling at Yakutsk array showed that true slope of radio emission LDF must be greater by 5-8 $\%$, which is required to take into account when estimating maximum depth of EAS X$_{max}$ using LDF form.

\section{Conclusion}

Extensive air showers radio emission measurements shows: a) developed algorithm clearly identifies showers with radio emission; b) radio emission slowly attenuates with a distance and registers  up to 1000 m from shower axis with energy more than 3$\cdot$10$^{17}$ eV; c) it can be used for X$_{max}$ estimation,  using LDF form as alternative for optic measurements [3].

{\bf Acknowledgements:} We express our thanks to the Ministry of Education and Science of the Russian Federation for the financial support of this study as part of the RFBR 12-02-31442 mol$\_$a.

\end{document}